# Catalytic effect of plasma in lowering the reduction temperature of $Fe_2O_3$

Jaemin Yoo[1], Dongkyu Lee[2], Jimo Lee[1], Taehyeong Kim[2], Hyungyu Jin[2], Gunsu S. Yun[1,3,4*]

[1]*Division of Advanced Nuclear Engineering, Pohang University of Science and Technology, Pohang, 37673, Republic of Korea*
[2]*Department of Mechanical Engineering, Pohang University of Science and Technology, Pohang, 37673, Republic of Korea*
[3]*Department of Physics, Pohang University of Science and Technology, Pohang, 37673, Republic of Korea*
[4]*Max Planck Center for Attosecond Science, Max Planck POSTECH/Korea Research Initiative, Pohang, 37673, Republic of Korea*

*Email address: gunsu@postech.ac.kr

**ABSTRACT**

Atmospheric pressure plasma (APP) generates highly reactive species that are useful for surface activations. We demonstrate a fast regeneration of iron oxides, that are popular catalysts in various industrial processes, using microwave-driven argon APP under ambient condition. The surface treatment of hematite powder by the APP with a small portion of hydrogen (0.5%) lowers the oxide's reduction temperature. A near-infrared laser is used for localized heating to control the surface temperature. Controlled experiments without plasma confirm the catalytic effect of the plasma. Raman, XRD, SEM, and XPS analyses show that the plasma treatment changed the chemical state of the hematite to that of magnetite without sintering.

**INTRODUCTION**

Capability of exchanging oxygen in metal oxides via thermochemical redox reactions has led to production of gas fuels like carbon monoxide, hydrogen, or syngas, as well as to storage of thermal energy in the matters.[1-3] The exchange of oxygen starts from the loss of lattice oxygen, promoting the evolution of oxygen and better absorption of heat for the matter. Subsequently, the reduced oxide is reoxidized by various oxidants like $CO_2$, $H_2O$, $O_2$ to produce gas fuels. Such thermochemical reactions can be improved when the material has a large extent of oxygen exchange. That is to say, materials having large redox capacity can produce more amount of gas fuel and store more energy.

Iron oxides are a promising system for redox catalysis due to the low third ionization potential compared to other non-stoichiometric metal oxides. Despite the potentially large redox capacity the iron oxide has, its rapid sintering and high reduction temperature have hindered practical use in fuel production or energy storage.[1,4] The high reduction temperature limits the selection of catalysts used in the reactor system, in which additional components for heat recovery are essential to enhance the system efficiency.[5-7] Plasma is expected to improve the efficiency as it can induce heat an intended area with plasma-self heating without heating the entire sample area compared to a conventional reactor system, and also lower the reduction temperature and shorten the reaction time.

Plasma has an exceptional advantage in making neutral species highly reactive for various chemical reactions. Enhancement of electro-chemical reactivity by plasma can occur under a wide range of gas pressure or temperature. Abundant radicals generated from plasma is known to increase the internal energy of the reacting species, and thus more reactions can occur as the activation barrier in the chemical reaction is lowered.[8]

Much attention has been given to atmospheric pressure plasma (APP) recently for its high potential of continuous operation in an open space without the need for a bulky chamber. While APP can be discharged with various power sources, ranging from DC to GHz, microwave is particularly known for its highly effective confinement of electrons combined with high displacement current, generating energetic electrons and promoting higher production rates of reactive species. [9-15] Some other characteristics of microwave-driven plasma include low voltage/power operation, long lifetime electrode, and low production of ozone. [10,11,16,17]

In this letter, we report a catalytic effect of plasma in lowering the temperature required for the reduction of hematite using a compact microwave-driven plasma source called a Coaxial Transmission Line Resonator (CTLR).[9] A very small amount of hydrogen, an intense reducing agent, and a near-infrared laser (808 nm) under continuous wave mode for localized heating are used to enhance the rate of reduction process. More details of experimental setup can be found in Supplement 1.

**RESULTS AND DISCUSSION**

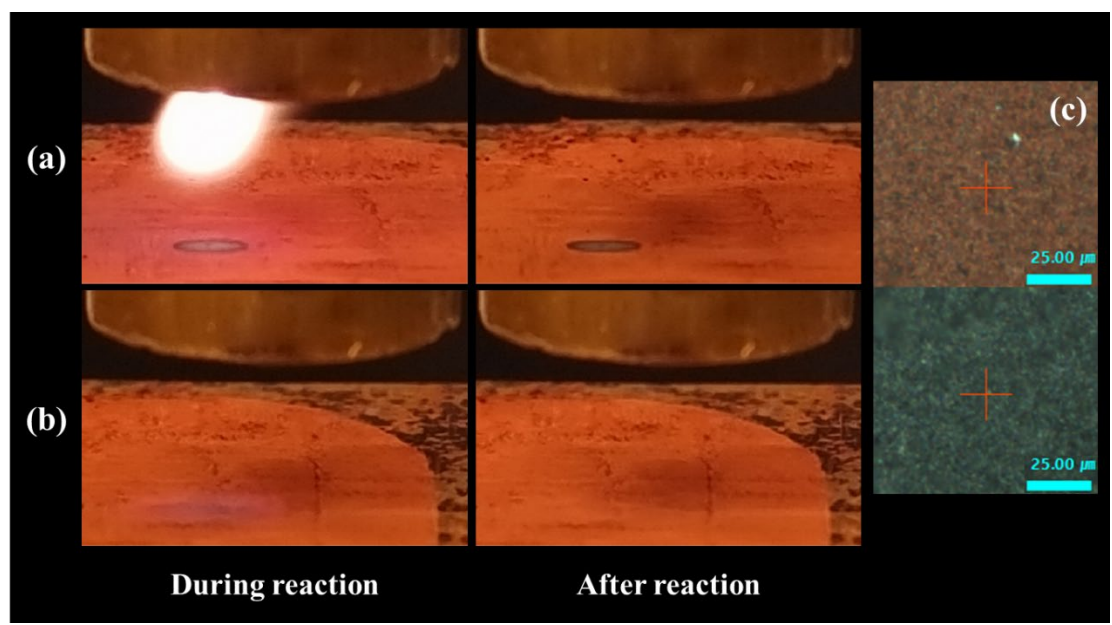

**FIGURE 1** Reaction processes of: (a) Both plasma and laser on and (b) Laser on only, and (c) microscopic images of untreated hematite (top) and treated hematite (bottom)

Figure 1(a) and 1(b) show two different combinations of plasma and laser treatments on the hematite surface at $T_{rad} \approx 630$ K, and 5-minute treatments are performed to ensure sufficient reaction. Under such conditions, dramatic color change on the surface is observed in 1(a) case, whereas no distinct change is detected with the laser only case. These results suggest the probable catalytic effect of plasma on hematite powder, that is lowering the reaction barrier of metal oxides. Three different, could be independent or dependent, factors are suggested: microwave heating effect, overabundant energetic electrons, and hydrogen reactive species. A few studies on thermochemical redox reactions via microwave heating have been reported. [18-20] It is reported that the direct radiation of microwave can promote the loss of lattice oxygen on metal oxides. The microwave-driven plasma introduced in this letter, however, is positioned far from the surface. The heating effect of microwave plasma is thus perceived to be negligible. Abundant electrons generated by the plasma are presumed to be the main factor of low reduction temperature as the electron plays a major role in the formation of polarons and oxygen vacancies in metal oxides.[18] As a small portion of hydrogen (0.5%) mixed with argon is used for the plasma discharge, reactive hydrogen species ($H_2^*$, $H^+$, H etc.) cannot be neglected. It is proposed that $H_2^*$ (ro-vibrationally excited hydrogen molecule) is the dominant radical in microwave-induced non-thermal hydrogen plasma resulting in the reduction process of metal oxides.[21-23]

The change in the hematite surface is diagnozed using an X-Ray Diffractometer (XRD) and a Raman spectrometer, a Scanning Electron Microscope (SEM), and an X-ray Photoelectron Spectrometer (XPS).

**Scanning Electron Microscope (SEM)**

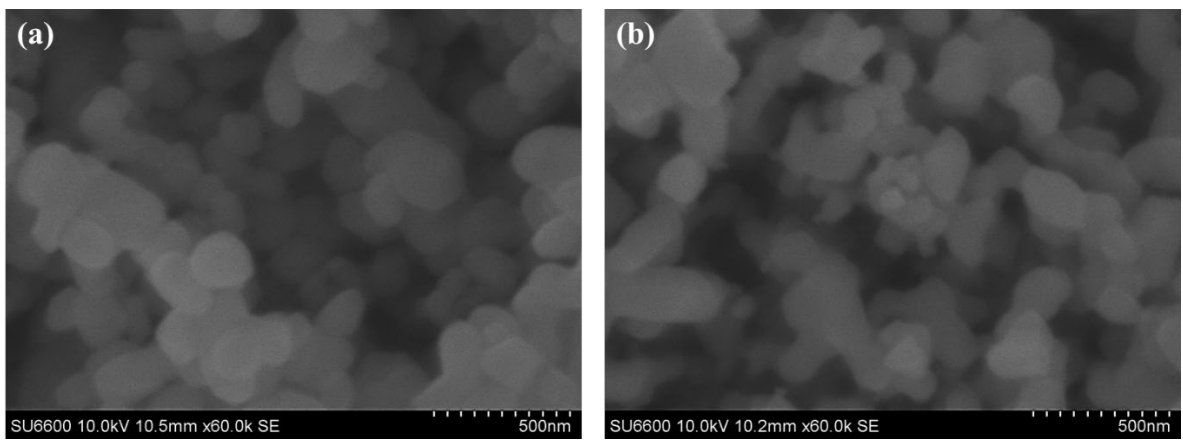

**Figure 2**  SEM images of (a) untreated hematite and (b) treated hematite for 5-minute.

We analyze the morphologies of hematite particle by SEM before and after the plasma treatment. The morphologies of samples are shown in figure 2. The untreated hematite particles have the size under 500nm, and the size of particle is maintained after the plasma treatment for 5-minute without evidence of sintering. Our finding would seem to imply that the plasma treatment is a suitable tool for reducing metal oxides.

**X-Ray Diffraction (XRD) pattern in samples**

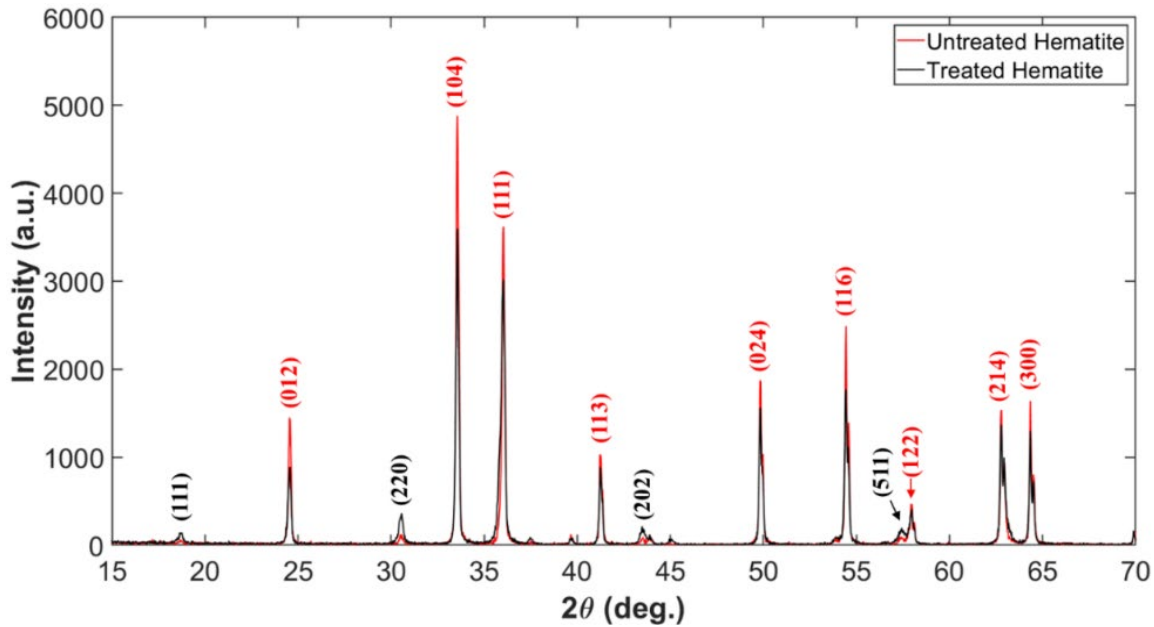

**FIGURE 3**  XRD patterns of untreated and treated hematite samples.

Figure 3 shows the diffraction patterns of untreated and treated hematite powder. It is note that the patterns contain both signals from the surface and the bulk of the sample as the XRD measures bulk diffraction patterns. Miller indices planes obtained from the Inorganic Crystal Structure Database (ICSD) are indicated. Miller indices denoted by (012), (104), (111), (113), (024), (116), (122), (214) and (300) are signals that appear in a hematite sample (ICSD code 01-080-5407 and 01-089-8103).[24] Miller indices denoted by (111), (220), (202) and (511) are signals that appear in a magnetite sample (ICSD code 01-071-6336 and 01-076-2949).[24,25] Figure 2 shows that the plasma treatment has changed the chemical state of hematite to that of magnetite. In addition, the weakening of the signal intensity of the treated hematite indicates that the treated sample has a state of the 2-phase mixture (hematite-magnetite). The observation of fine shift in main signals, (104), and (111), is an evidence of a phase shift.

**Raman spectroscopic and X-ray Photoelectron spectroscopic analysis**

We use Raman spectroscopy for cross-validation with XRD results, and X-ray Photoelectron Spectroscopy (XPS) to analyze the chemical state of the sample surfaces. Figure 4(a) presents the Raman spectra of untreated and treated hematite. The observed peaks at $A_{1g}$ and $E_g$ correspond to the phonon vibrational mode of hematite, and the continuum signal appearing near 1318 cm$^{-1}$ corresponds to the two-magnon scattering of hematite.[26,27] The $A_{1g}$ peak near 668 cm$^{-1}$ corresponds to the vibrational mode of magnetite.[28,29] The partially weakened hematite peaks and $A_{1g}$ of magnetite peak show that the hematite has a state of the two-phase mixture which is consistent with the results of XRD patterns.

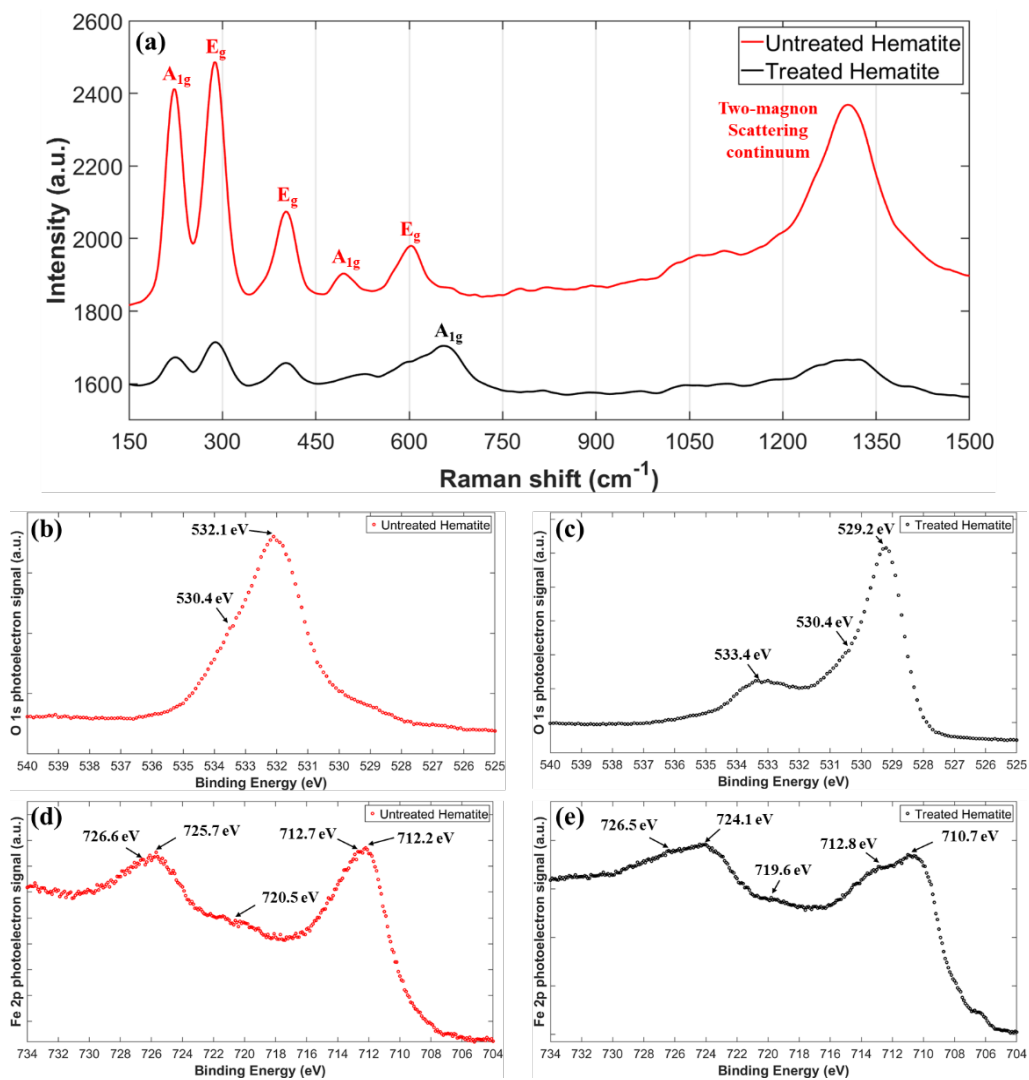

**FIGURE 4** (a) Raman and (b)-(e) photoelectron spectra collected from the untreated and treated hematite. (b) O1s for untreated hematite, and (c) O 1s for the treated hematite. (d) Fe 2p for the untreated, and (e) Fe 2p for the treated hematite.

The surface electronic state of the samples is analyzed by XPS. While the untreated sample shows the main peak of 532.1 eV, the treated sample shows a peak at higher energy around 533.4 eV which is attributed to the surface oxygen defect species as shown in figure 4(b) and (c).[30-32] Figure 4(d) and (e) show the representative Fe $2p_{3/2}$ (710.7 eV) and Fe $2p_{1/2}$ (724.1 eV) peaks of untreated and treated hematite.[33,34] Compared to the untreated hematite, the Fe 2p peaks of treated hematite increment in the lower energy around 710.7 eV and 724.1 eV indicating that some of $Fe^{3+}$ are reduced by creating oxygen vacancies.

# CONCLUSION

Here we proposed a novel technique to control the surface chemical state of hematite using microwave-driven atmospheric pressure plasma. SEM, XRD, Raman, and XPS analysis show that the chemical state of the hematite powder was partially changed to that of magnetite by the plasma with laser-heating under ambient pressure, and the rate of reduction can be accelerated. The catalytic effect of microwave-driven plasma is confirmed by the unaffected surface of the laser-treated sample.

The surface temperature is a crucial parameter in the plasma-metal oxide reaction. The main advantage of the laser as a heating source is the acceleration of the reaction rate by compensating for the lack of local heating ability of plasma. The combination of the high reactivity of microwave plasma and fast surface heating by laser has several advantages in terms of energy efficiency and time scale of the desired reaction as compared to the laser alone case. Local surface heating of laser can minimize undesired reactions as well as prevent the destruction of the target. Laser also proves to be more productive in terms of energy efficiency than the convection heating by avoiding the massive heat transfer to a bulk region and by minimizing the cooling effect caused by an external parameter like gas flow.

The reduction of hematite using plasma shows the state of a partially reduced 2-phase mixture rather than the completely reduced state as metallic iron (Fe). It appears that the particle size of hematite powder is maintained after the treatment, and the plasma lowers the temperature at which reduction begins to occur. Such thermodynamic advantages of the plasma treatment can be utilized in hydrogen production.

It is noted that the proposed study has one limitation on temperature measurement. The infrared imaging camera indirectly measures the surface temperature, and the change in surface emissivity during the reduction reaction of hematite to magnetite is also not considered. Thus, there is a limit to the thermodynamic analysis of a chemical reaction with temperature dependence. Developing a tool to measure the surface temperature of the plasma-metal oxide interface is left as future work.

We have presented three probable hypotheses of the low reduction temperature; effects of microwave heating, surface charging of electrons, and hydrogen radicals. Further studies of their roles are subjected to be carried out in future work.

**Acknowledgements**

This work is partially supported by the BK21+ program funded by the Ministry of Education, Republic of Korea, the POSTECH-Samsung Education Program, and the KOREA HYDRO & NUCLEAR POWER CO., LTD (No. 2019-TECH-17)

# Supplement 1

**EXPERIMENTAL SETUP**

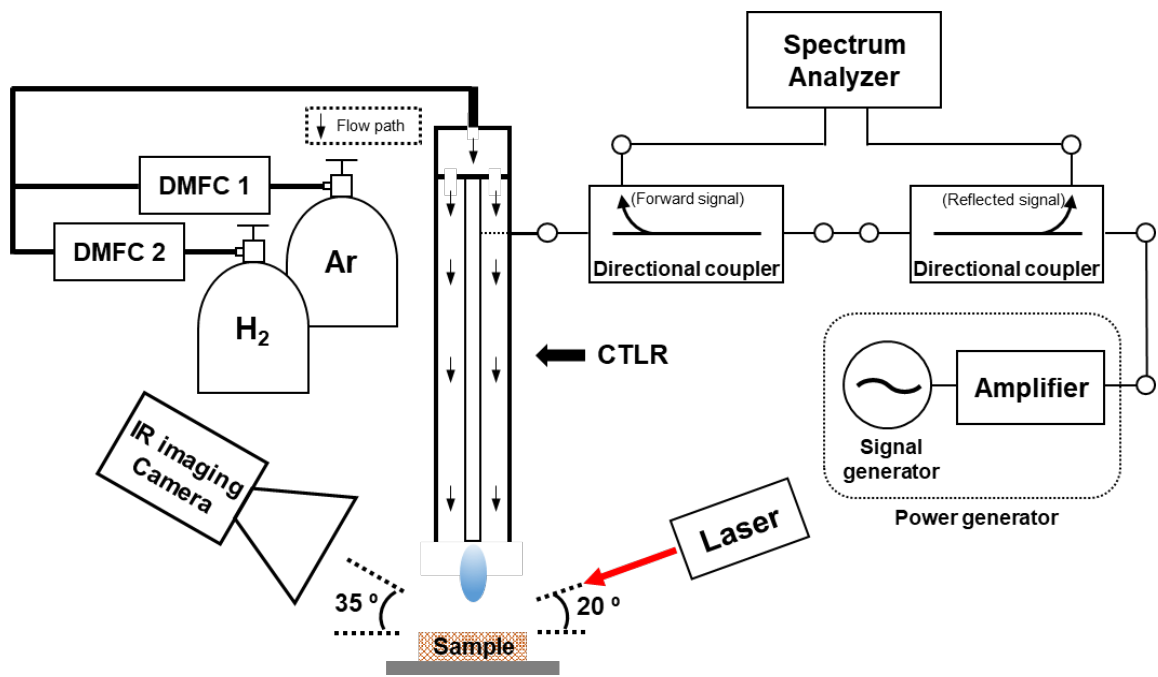

**FIGURE S1.**  Experimental schematic diagram.

Figure S1 shows the experimental schematic diagram of the microwave CTLR plasma system combined with a laser. We use argon as a carrier gas and hydrogen as a feed gas to boost the generation of radical hydrogen species. The system consists of the CTLR electrode, digital mass flow controllers (VIC-D200 Series, MFC KOREA), and a microwave source. In the microwave source, a power generator (2.45GHz ISM BAND SSPA, KRF) with a signal generator and an amplifier is used to drive a plamsa. The net input power coupled to the plasma is measured using a spectrum analyzer (E4408B, Agilent) with two directional couplers (ZUDC20-183+, Mini-Circuits) guiding the forward and reflected signals. The compact laser operating in a continuous wave with 808 nm and 1.5 W controls the surface temperature of a sample, and the laser path angle to the sample plane is 20 degrees.

The sample used in the experiments is hematite in the form of powder (particle size < 5 μm, purity 99 %) (310050-500G, Sigma-Aldrich). We prepare the sample by compacting the powder in an copper plate with a shallow recess in the middle to prevent its dispersion by the gas flow. During the treatment of the sample, we monitor the surface temperature of the sample using an IR imaging camera (P620, FLIR) with a notch filter to block the laser light. The camera angle to the sample plane is 35 degrees. Since the emissivity data for hematite powder is not available, we set the emissivity of 0.95 as a reasonable value.[46, 47] After the treatment, we analyze the oxidation state of the sample using an X-Ray diffractometer (RINT 2000, Rigaku), a Raman spectrometer (FEX-MD, NOST), Scanning Electron Microscope (SU6600 ,Hitachi), and X-ray Photoelectron Spectrometer (ESCALAB 250, Thermo Scientific).

The flow rates of individual gases are critical in obtaining high-density plasma while maintaining the stability of the plasma plume for a given microwave input power. For instance, too little argon flow will result in a small plasma plume size, while too much argon flow will result in an unsteady turbulent flow. On the other hand, too much hydrogen flow will lead to a small plasma volume. For a nominal input power level of 15 W, we achieve a stable plasma plume with an argon flow rate of about 2 standard liter per minute (SLM), and an hydrogen flow rate of about 10 standard cubic centimeters per minute (SCCM).

While maintaining the gas flow of argon and hydrogen through the center of electrode, the surface of the hematite is confirmed to reduce from a radiation temperature ($T_{rad}$) of about 570 K in plasma and laser on condition. The laser only condition, however, required a much higher $T_{rad}$ of about 720 K for the surface to be reduced. Interestingly, this result is significant evidence that the plasma lowers the reaction barrier.